# A high performance scientific cloud computing environment for materials simulations

K. Jorissen*, F.D. Vila, and J.J. Rehr

Department of Physics, University of Washington, Seattle, WA 98195, USA
* corresponding author

## Abstract

*We describe the development of a scientific cloud computing (SCC) platform that offers high performance computation capability.  The platform consists of a scientific virtual machine prototype containing a UNIX operating system and several materials science codes, together with essential interface tools (an SCC toolset) that offers functionality comparable to local compute clusters.  In particular, our SCC toolset provides automatic creation of virtual clusters for parallel computing, including tools for execution and monitoring performance, as well as efficient I/O utilities that enable seamless connections to and from the cloud. Our SCC platform is optimized for the Amazon Elastic Compute Cloud (EC2).  We present benchmarks for prototypical scientific applications and demonstrate performance comparable to local compute clusters. To facilitate code execution and provide user-friendly access, we have also integrated cloud computing capability in a JAVA-based GUI.  Our SCC platform may be an alternative to traditional HPC resources for materials science or quantum chemistry applications.*

## 1. Introduction

Cloud Computing (CC) is a computational paradigm in which dynamically scalable, virtualized resources are provided as a service over the internet.[1-4]  This paradigm has seen remarkable advances over the last few years, especially with the emergence of several commercial cloud services that take advantage of economies of scale.[5-9]  While many commercial applications have quickly embraced CC developments, scientists have been slower to exploit the possibilities of a CC environment.  Scientists are not new to shared computing resources, such as Beowulf clusters, which are often needed for modern condensed matter and materials science simulations.  Also cloud-like resources such as Grid Computing and CONDOR clusters have been useful for some scientific applications. However these latter resources are typically loosely coupled, inhomogeneous, and geographically dispersed, and not well suited for the high performance computing (HPC) demands of many scientific codes. Recently, dedicated scientific cloud test beds have begun to be explored by national research facilities such as NERSC[10] and the NEON network[11]. Additionally, a number of studies have explored the concept, feasibility, or cost-effectiveness of cloud computing for research.[12-16] There have been commercial and community efforts to develop tools that make access to cloud resources easier. Notably the StarCluster project[17] provides a utility for managing general purpose computing clusters on EC2. However, significant further developments were needed to create a platform for materials simulations that meets all the particular needs of HP scientific computing without requiring further configuration, and is accessible not only to system administrators but also to general users. Furthermore the questions of cost-effectiveness and performance have not been conclusively answered and need to be addressed for each type of scientific cloud application. In particular, concerns about CC performance are strong in the materials science community. Here we demonstrate that with the advent of UNIX-based HPC cloud resources such as the Amazon EC2 and the 2$^{nd}$ generation cloud cluster tools described below, there is now considerable potential for Scientific Cloud Computing (SCC). In particular, we show that SCC is especially appropriate for materials science and quantum-chemistry simulations, which tend to be dominated by computational performance rather than data transfer and storage.

Recently we established proof of principle for the feasibility of SCC for some prototypical scientific applications. [18] In particular, we created an AMI (Amazon Machine Image) containing parallel codes, and a first generation set of tools to create and control clusters of virtual machines on the Amazon Web Services (AWS) Elastic Compute Cloud (EC2) (Fig. 1). These tools are shell scripts that can be run from the command line on *NIX systems. Benchmarks in this environment showed that a parallelized scientific code with modest requirements in terms of memory and network speed, yielded similar performance on a virtual EC2 cluster as on a local physical cluster.[18] However, at the time the capabilities of the EC2 were limited by the high latency and low bandwidth of the cluster interconnects. Thus in the present work we describe a virtual SCC platform that takes advantage of current HPC cloud resources and demonstrates scalability and performance comparable to local compute clusters.

This goal is accomplished in several steps: 1) We briefly describe the elements of our SCC AMI, i.e., the operating system and HPC scientific applications included in the improved AMI. (From here on, we will generally use the AWS specific terminology: "AMI" for a virtual machine image, and "EC2" for the Cloud.) 2) We describe a 2$^{nd}$ generation SCC toolset, consisting of `bash` scripts, that makes the EC2 cloud perform virtually like a local HPC UNIX cluster, and we verify its improved performance. 3) We present benchmarks for parallel performance of HPC scientific applications, focusing in particular on the intranet performance. 4) Finally, in order to facilitate access to the EC2, we have developed a graphical user interface (GUI) that controls execution and I/O for a prototypical application.

We focus here primarily on scientific applications in condensed matter physics, materials science, and quantum-chemistry. Most applications in those fields have a workflow and coding characteristics that rely on computational performance, rather than data-managing capability. These special features in turn influence the choice of cloud paradigm. A key feature of such applications is their simple control and data workflows. Typical simulations involve a set of small input files of a few KB that define the parameters for the run; a series of computationally intensive steps; and the production of a set of small to medium size output files, typically ranging from 1-100MB. Thus, typical materials-science simulations differ from data-driven cloud applications that can take advantage of software frameworks such as MapReduce[19] or Hadoop[20]. Nevertheless, given that there is very little communication to and from the cloud and that data transfer can be a substantial component of cloud computing costs, materials science applications tend to be relatively cost-effective data wise. Another key trait of many scientific applications is their legacy character. For instance, many codes are written in FORTRAN and make extensive use of the Message Passing Interface (MPI) for parallelization. Consequently, the deployment of these applications to an Infrastructure-as-a-Service (IaaS) cloud environment such as EC2 is highly advantageous. IaaS environments can be configured to match the non-standard requirements of most legacy applications and offer the added advantage of providing traditional homogenous environments that are highly desirable for both users and developers. In contrast, Platform-as-a-Service (PaaS) environments such as Microsoft's Windows Azure Platform[21] require major changes in software structure and, at least at present, are not able to accommodate MPI.

For applications that are highly data-intensive, cloud computing may not currently be competitive. While the per-GB cost of Simple Storage System (S3) storage[22] can be comparable to that of local solutions (e.g. [23]), and has the advantage of elasticity (one only pays for the storage or transfers needed at a given time), cloud storage can have considerable disadvantages. For example, local tests showed that data transfer to and from S3 peaked at 0.032Gbps [24] while a local service of comparable cost delivered 10Gbps. If bandwidth is essential, this objection alone makes cloud storage unviable. Furthermore, data transfer to and from S3 is charged per TB. Therefore, such communication is slow and costly for users who frequently need to move large amounts of data back and forth. For data that does not need to be accessed or moved much, tape archives are currently still a cheaper solution for archival purposes compared to S3 storage.[23] For decades, supercomputer centers have offered safe, cheap archive services for large datasets that almost never change, coupled with finite amounts of spinning (expensive) scratch disk. The researcher typically uses the archive as the main data store, staging data onto spinning disk only when it is needed for calculation. Once the researcher is finished computing, any new data is moved back to the archive. This model is currently the most cost-effective way of meeting the storage needs of scientists in an elastic computing environment and there is no competitive storage mode in the AWS Cloud. [24] While these objections are not important for the typical materials science applications discussed in this paper, they would have to be considered carefully for more data-intensive work.

Although the principles outlined above are applicable to other cloud computing environments with comparable capabilities, the developments presented here are currently designed specifically for the EC2. In addition to the IaaS argument given above, this gives us the advantage of a major cloud provider with large capacity and a large user community, whose application programming interface (API) is somewhat of a standard. Finally, EC2 offers virtual HPC resources, which we discuss in Section 4. The tools we have developed, however, are intended to be generic and could be easily modified to use other cloud providers, or serve other applications where HPC resources are essential. Many further developments such as the inclusion of a wider variety of scientific codes with improved interfaces are now straightforward. Our developments have led to a functional, beta-stage HPC SCC platform with the potential to make high performance scientific computing available to those who lack expertise and/or access to traditional HPC resources.

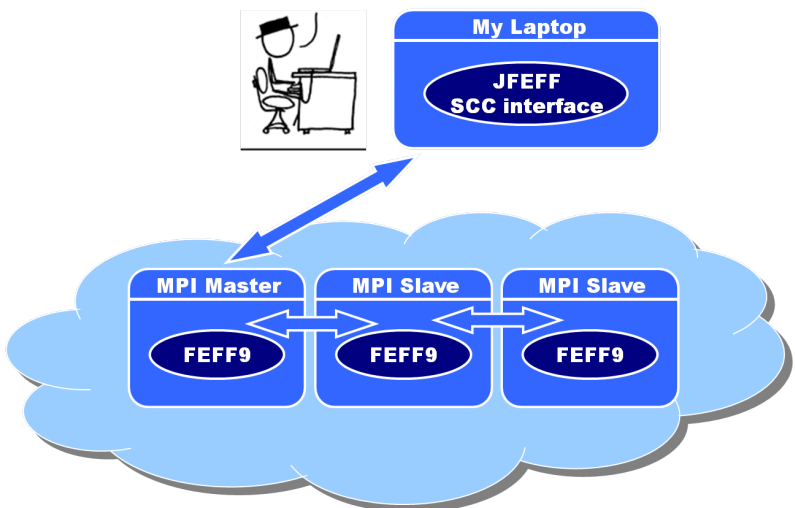

**Figure 1** Illustration of the cloud environment. The user runs an interface on a local machine. This interface could be a GUI (e.g. JFEFF), or our command line toolset. The interface creates a virtual cluster in the cloud and controls the execution of jobs on this cluster. When finished, the interface retrieves the results and terminates the virtual cluster.

## 2. The Scientific Cloud Computing Machine Image

In this section we briefly describe the scientific cloud computing AMI. This virtual machine image serves as a blueprint for a cloud instance specifically configured for parallel, HPC scientific computing applications. In a nutshell it is a minimal 64-bit

Fedora 13 LINUX distribution that is enhanced with tools typically needed for scientific computing, such as Fortran 95 and C++ compilers, numerical libraries (e.g. BLAS, ScaLAPACK), MPI (1.4.3), PBS, PVFS, etc. Our 2$^{nd}$ generation SCC AMI has several improvements over the original prototype[18]: This AMI is bundled with several widely used materials science codes that typically demand HPC capabilities. These include electronic structure codes (ABINIT[25], Quantum ESPRESSO[26], Car-Parrinello[27], and WIEN2k [19]), and excited-state codes (AI2NBSE[28], Exc!ting[29], FEFF9[17], OCEAN[30], RT-SIESTA[31]), and quantum chemistry codes (NWChem[32]), although here we only present benchmarks for FEFF9 and WIEN2k. The 64-bit LINUX operating system is better suited for scientific computing than its 32-bit predecessor in the prototype [16]. This machine image is now stored on Amazon's Elastic Block Storage (EBS)[33] system, rather than the older S3[22] infrastructure, leading to a reduction in instance boot times of 20-50% (see Section 2.4). Depending on needs, the new SCC AMI can be loaded onto different "instance types". Slower but cheaper instances can be used for simple tasks, while higher performance workhorses are available for more demanding calculations. In particular, in Section 4.2 we discuss the new EC2 "Cluster Instances"[34], which are very important for HPC.

## 3. The Scientific Cloud Computing toolset

### *3.1. Functionality*

Our 2$^{nd}$ generation SCC toolset consists of a handful of `bash` scripts that run on a local machine (e.g., a laptop PC or UNIX desktop), and are used to control the virtual SCC environment. The functionality of the toolset is twofold: First, it transforms a group of instances created by EC2 based on our AMI, into an interconnected cluster that functions as a virtual parallel computing platform. How this is done is described in detail below for the "ec2-clust-launch" script. Second, the toolset is a wrapper for the EC2 API, replacing cumbersome API calls by much more user-friendly calls that store many settings in the environment and in configuration files to keep the user from having to manage them manually. They function as an intermediary layer between the user and the EC2 API[35]. For example, a user could open an SSH session on an existing EC2 instance by hand from a command-line terminal by entering a rather complicated, session-dependent command

```
ssh -i/home/user/.ec2_clust/.ec2_clust_info.7729.r-de70cdb7/key_
pair_user.pem user@ec2-72-44-53-27.compute-1.amazonaws.com
```

Alternatively, using the SCC toolset script, the same task only requires

```
ec2-clust-connect
```

The toolset also simplifies the use of applications within the cluster by providing scripts for launching and monitoring the load of different tasks.

*3.2. Description of the tools*

Table 1 lists the currently available commands in the SCC toolset. All these commands are installed on the user's local machine and act remotely on the virtual cluster. Moreover, some of these tools also have counterparts installed on the AMI that can be used to monitor the execution from within the cluster. The functionality of the tools is briefly summarized below, together with their LINUX equivalents.

**Table 1**
SCC toolset commands to launch and interact with virtual EC2 clusters and their LINUX counterparts.

| Name | Function | Analog |
| --- | --- | --- |
| `ec2-clust-launch` *N* | Launch cluster with *N* instances | `boot` |
| `ec2-clust-connect` | Connect to a cluster | `ssh` |
| `ec2-clust-put` | Transfer data to a cluster | `scp` |
| `ec2-clust-get` | Transfer data from a cluster | `scp` |
| `ec2-clust-list` | List running clusters | `top` |
| `ec2-clust-terminate` | Terminate a running cluster | `shutdown` |
| `ec2-clust-run` | Start job on a cluster | `run` |
| `ec2-clust-usage` | Monitor CPU usage in cluster | `top` |
| `ec2-clust-load` | Monitor load in cluster | `loadavg` |

```
ec2-clust-launch –n N [–c Name] [–m MachineType] [–t
InstanceType] [–e EphStorage]
```

The `ec2-clust-launch` script is the most important tool in the set: It performs all the tasks needed to launch and configure a cluster of *N* instances on the EC2. Optionally, the *MachineType* (i.e. AMI) and the *InstanceType* can be selected. AWS currently offers about a dozen *InstanceTypes* of varying CPU, memory, and network capabilities.[36] Schematically the ec2-clust-launch script performs the following tasks, as summarized from the comments within the script:

```
#!/bin/sh
### Create a cluster of CLUST_NINS instances
# Get the general configuration information
# Check if the EC2_HOME is set and set all the derived variables
# Check and set the location of the cluster tools
# To avoid launch problems check for the presence of a lock
```

```
# Create a lock for this process
# Check if we have a cluster list
# Get the total number of instances
# Set the default cluster name (use current process id)
# Set the default machine type
# Process input options
# Check that the PK (private key) and CERT files are available
# Set the cluster index
# Load the appropriate machine profile
# Create an EC2 placement group if requesting cluster instances
# Launch instances on EC2
# Get reservation ID and list of instance IDs
# Get the instance rank indices
# Save name and reservation ID in cluster list
# Release the lock
# Make a directory to hold the cluster information
# Manage the certificates that are used to access the cluster
# Monitor instances until all the information we need is available
# Make directory that will be used to store info to transfer
# Initialize setup script in transfer directory
# Get public and private DNS names
# Set the head instance public DNS name
# Create a list of the internal EC2 addresses
# Save information about the cluster to .ec2_clust_config file
# Create a hosts file and mpi hostfile for all the cluster instances
# Copy hosts files to directory to transfer and add to setup script
# Copy monitor tools to directory to transfer, add to setup script
# Set up ephemeral storage on cluster instances
# Point SCRATCH file system to ephemeral volume
# Add shared dir creation to setup script
# Create the exports file for the head instance
# Copy exports file to directory to transfer and add to setup script
# Add nfs config reload to setup script
# Add fstab update and mount to setup script
# Copy user certificates to directory to transfer
# Add user certificate setup to setup script
# Compress the info storage directory
# Make sure the keys are ready on the other side
# Transfer all files at once but don't launch more processes than
permitted by the OS
# Run setup on all nodes
# Optionally give the head node a head start so it can get the nfs
exports ready by the time the nodes want to mount them
# Do cleanup locally but save cluster information
# Print out setup timing info
```

We now discuss this set of operations further. Each of the *N* nodes is a clone of the selected AMI, with all its preinstalled software and data, running on virtualized hardware determined by the *InstanceType* (e.g., "High CPU, 8 cores"). When the *N* instances have booted in EC2, the launch script performs setup tasks that transform the *N* individual machines into an *N*-node cluster that functions like a traditional LINUX Beowulf cluster. The tasks mentioned above include mounting an NFS partition and creating appropriate "/etc/hosts" files on all nodes, and configuring passwordless ssh access between nodes, all of which are requirements for many parallel scientific codes. One node is designated master node. This node generally distributes MPI jobs to the other nodes and makes a 'working directory' partition available over the local network. The script also sets up a user account other than root for users to run the scientific codes provided in the AMIs. It is useful to tag the cluster with a *Name* (-c argument), especially if one intends to run several clusters at the same time. Certain *InstanceTypes* can create additional ephemeral data volumes up to about 2TB for storage intensive calculations (-e argument). The `ec2-clust-launch` command creates a temporary folder on the local control computer to store information about the cluster. This information includes identifiers and internal and external IP addresses for each of the instances comprising the cluster. The other scripts in the toolset access this information when they need to communicate with the cluster. User-related information, e.g. identifiers for the user's AWS account, is stored in environment variables.

`ec2-clust-connect [-c Name]`

Opens a ssh session on the *Name* cluster, or on the most recently launched cluster if no argument is given. The script logs in with the user account created by `ec2-clust-launch`, instead of the default root access offered by AWS.

`ec2-clust-connect-root [-c Name]`

Opens a ssh session on the *Name* cluster and logs in as root. This is required only for developers, not for users running a calculation, unless runtime changes in configuration are needed.

`ec2-clust-put [-c Name] localfile remotefile`

Copies the file *localfile* on the local machine to the file *remotefile* on the master node of the *Name* cluster (or the most recent cluster if none is specified). If *localfile* is a directory it will be copied recursively. The master node has a shared working directory that all other nodes can access.

`ec2-clust-get [-c Name] remotefile localfile`

Copies the file *remotefile* on the head node of the *Name* cluster (or the most recent cluster if none is specified) to the file *localfile* on the local machine. If *remotefile* is a directory it will be copied recursively. The master node has a shared working directory that all other nodes can access.

`ec2-clust-list`

Lists all active clusters. Each cluster is identified by a *Name*, its AWS reservation ID, and an index number.

```
ec2-clust-terminate [-c Name]
```

Terminates all *N* instances comprising the cloud cluster *Name*, and cleans up the configuration files containing the specifics of the cluster on the local machine. The cluster cannot be restarted; all desired data should be retrieved before running the 'terminate' script. If no cluster is specified, the most recent one will be terminated.

```
ec2-clust-run –e Task [-c Name] [-t]
```
This tool connects to cluster *Name* (or the most recent cluster if none is specified) and executes a job there. Currently, *Task* can be WIEN2k or FEFF9. The tool loads a profile describing the selected *Task*. It scans the working directory for required input files and copies them to the cloud cluster *Name*. It then instructs the cluster to execute the task on all its nodes. It periodically connects to check for *Task* specific error files or successful termination. It copies relevant output files back to the local working directory, and terminates the cluster after completion if the *–t* flag is given.

```
ec2-clust-usage [-c Name]
```

Reports current CPU and memory usage for all nodes in cluster *Name*. This command can be executed either from within the cluster or from outside it, where the *–c* option is not required.

```
ec2-clust-load [-c Name]
```

Reports the 1 min, 5 min, and 15 min average load for all nodes in cluster *Name*. As in the `ec2-clust-usage` case, this command can be executed either from within the cluster or from outside it, where the *–c* option is not required.

### 3.3. toolset System Requirements

The present version of the SCC toolset has the following system and software requirements: the Java EC2 API[35], the Java runtime environment (RTE), and a *NIX environment with Bash and Secure shells. Thus, the toolset can be installed under many common operating systems, including Mac OS and, using Cygwin[37] on MS Windows. In addition to these software requirements, the user needs a valid Amazon AWS account and appropriate security credentials, including a ssh key pair for the AWS account.

### 3.4. Speed and Security

A drawback of our original toolset[18] was its limitation to serial flow when launching clusters, as the launch script configured cloud nodes one at a time. This lead to cluster boot times that increased linearly with the number of nodes. It took about 12 minutes to prepare a 16 node cluster using the most lightweight version of our AMI (see "Cluster Setup (v1)" and "S3 backed (v1)" in Fig. 2). Clearly, such setup times are not acceptable for HPC clusters using hundreds of nodes.

To solve this issue in the 2nd generation toolset (labeled "v2" in Fig. 2), we have parallelized all setup tasks insofar as security is not compromised. That is, the local machine sends simultaneously to all nodes a small file containing configuration data and instructions, and then instructs each node to perform its setup tasks simultaneously with its peers. Separate transfers to each machine are necessary for security reasons, since we do not want to send the user login credentials at boot up time. Consequently the resulting setup time is now roughly independent of cluster size, as demonstrated by the "v2" results shown in Fig. 2. A more modest but still significant speedup is obtained by switching AMI storage from S3- to EBS-backed[33], so that the AWS system can copy blueprints into actual virtual machines more quickly. For large clusters consisting on the order of 50-250 nodes, our cluster setup usually remains equally fast, though we occasionally have to wait an exceptionally long time for a few of the instances (e.g. 2 out of 256) to boot in AWS.

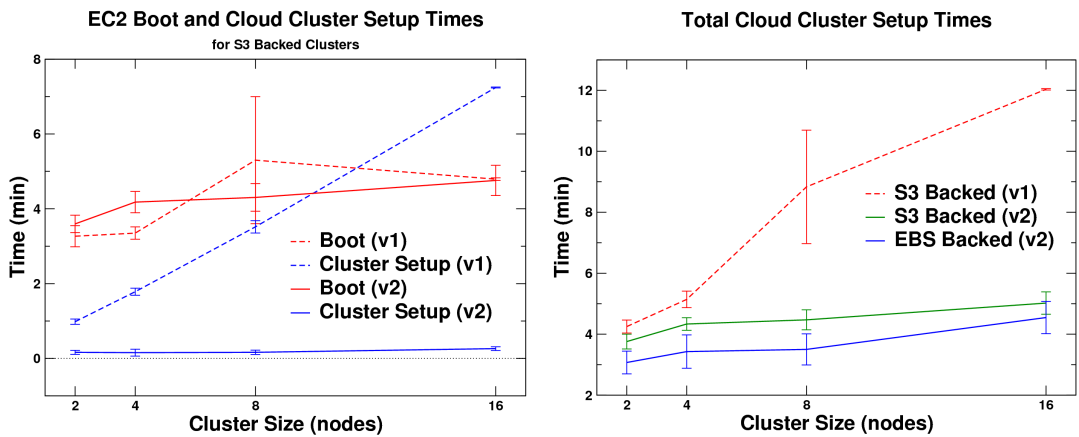

**Figure 2** EC2 cluster setup time using the 1st ("v1") and 2nd ("v2") generation toolset `ec2-clust-launch` script, as a function of the number of nodes in the cloud cluster. Right panel: total runtime of the launch script, including boot time of the EC2 nodes and cluster setup tasks. Left panel: boot time and cluster setup time shown separately for S3 backed clusters. During "Boot" we wait for Amazon EC2 to ready the nodes we have requested. During "Cluster Setup" we configure the nodes to form a SCC cluster.

## 4. Benchmarking the Scientific Cloud Computing Platform

To evaluate the capabilities of our SCC platform for practical applications, we have carried out performance benchmarks for two widely used materials science codes: FEFF9 [38,39] and WIEN2k[40,41]. Each has an active user base of over a thousand research groups in physics, chemistry, materials science, and biophysics. Both codes now serve as standard references, as evidenced by applications to a wide class of materials and material properties.[38-41] FEFF9 is an ab initio self-consistent multiple-scattering code for simultaneous calculations of excitation spectra and electronic structure.  The approach is based on a real-space Green's function formalism for calculations of x-ray and electron spectra, including x-ray absorption (XAS), extended x-ray absorption fine structure (EXAFS), electron energy loss spectroscopy (EELS), etc.  WIEN2k yields electronic structure calculations of periodic solids based on density functional theory (DFT), and uses the full-potential (linearized) augmented plane-wave ((L)APW) + local orbitals (lo) method. Like FEFF9, WIEN2k is a relativistic all-electron approach, and yields energy bands and phonons, as well as XAS and EELS, optical properties, etc. We do not discuss licensing issues here[42]; however, we have focused on codes which can be licensed to run on the EC2

The current trend in HPC is to distribute tasks more efficiently over a large number of cores, rather than exploiting faster CPUs. HPC codes are now developed accordingly. Thus HPC performance often hinges on fast communication between CPUs within a cluster. Heretofore, CC has been associated with lower intranet bandwidth and higher latency times than a dedicated local parallel cluster. This identifies one of the main concerns regarding the feasibility of High Performance Scientific Cloud Computing (HP-SCC): Are the intranet capabilities of cloud providers good enough to support HP-SCC?  As previously demonstrated[18], virtualization itself does not noticeably degrade performance, but massive numbers of instances are housed in vast hardware farms, where they have to share the network with many other instances, some of which may not even be in the same hub. We now show that these concerns are not always warranted, e.g. on the newer HPC instances.

In particular we compare results obtained on a virtual EC2 cluster to results on a local Beowulf cluster. This local UNIX cluster consists of AMD Opteron nodes with 16 cores each at 1.8GHz; 32GB memory; a 64bit platform; and connected with a 20Gbps Infiniband internal network.  Infiniband is currently the gold standard for networking, and typically has higher bandwidth and lower latency than Gigabit Ethernet networks.  For the cloud cluster, we consider two different types: The first, labeled 'Regular', consists of instances with 8 virtual cores of about 2.5GHz; 7 GB memory; a 64bit platform; and "high" network capacity ("high" being a qualitative assessment made by AWS). The second, labeled 'HPC' (i.e., 'Cluster Instance' in the AWS documentation), consists of instances that have 8 virtual cores at 2.93-3.33GHz; 23GB memory; a 64bit platform; and dedicated 10Gbps Ethernet interconnects. This latter instance type was recently introduced by AWS with HPC in mind.[34] It is the only instance type with a guaranteed, quantitative network speed. Indeed, it is somewhat less "virtualized" than the other instances, as it is never shared with other AWS users, and some specifications of the underlying hardware are available; e.g., each instance contains 2 Intel Xeon X5570, quad-core "Nehalem" CPUs.

We measure performance by the speedup ratio, defined as the time taken to run the same calculation on a single core divided by the time taken on N cores. The proximity of the slope of the resulting curve to a 1:1 ratio (perfect scaling) quantifies the degree of parallelization of the code and the quality of the network, which can degrade performance if it is not able to keep up with the code in shifting data between cores.

### *4.1. Loose coupling – the FEFF9 code*

The FEFF9 code[38] is naturally parallelized. The reason is that for calculations of x-ray and related spectra, nearly independent calculations have to be performed on an energy grid, and it is trivial to distribute these tasks over an array of processors using MPI. There is then very little need for communication between these parallel processes, except for I/O at the very end of the calculation, so we expect the FEFF9 code to be relatively insensitive to network performance.

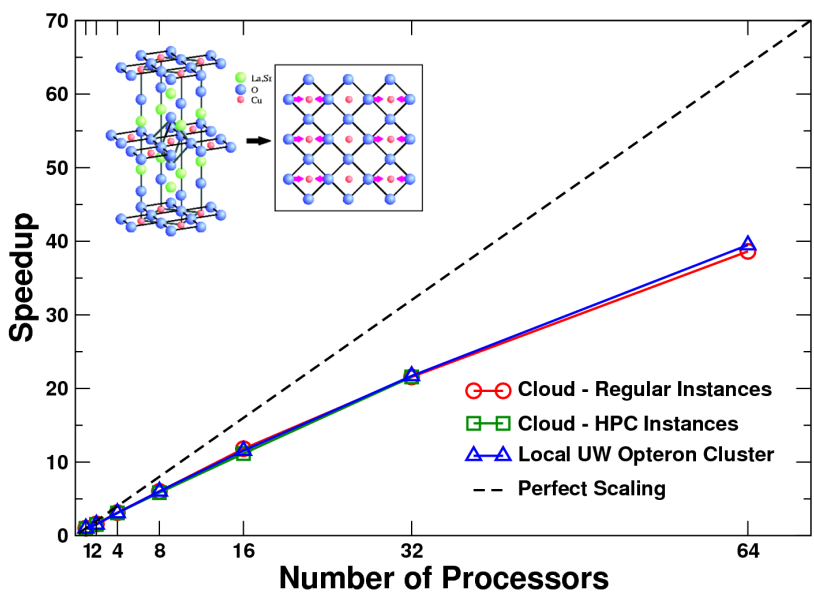

**Figure 3** Speedup of a FEFF9 calculation of the XAS of LSCO. The diagonal represents perfect scaling, where N processors finish the task N times faster than 1 processor. Performance on three different platforms is shown: a physical cluster ("Local UW Opteron Cluster") and two virtual clusters using "Regular" and "HPC Instances". The inset shows the structure of the LSCO high temperature superconductor.

Fig. 3 shows the speedup achieved by increasing the number of processors dedicated to a FEFF9 calculation of the X-ray Absorption Spectrum of a lanthanum strontium copper oxide (LSCO) high temperature superconductor on the physical Opteron cluster equipped with Infiniband, compared to two different virtual EC2 cloud clusters: one consisting of so-called "Regular" instances, which have network capabilities roughly equivalent to 100 Mbps Ethernet or worse; and so-called "HPC instances", which have dedicated 10 Gbps Ethernet connections and ought to deliver strong network performance. The FEFF9 code scales the same on all three architectures. Its parallelization has such a light footprint as to be insensitive to network performance. This confirms our earlier tests.[18]

*4.2. Tight coupling – the WIEN2k code*

The WIEN2k code has a more tightly coupled structure than FEFF9. In particular the Hamiltonian for a periodic system must be diagonalized on a grid of *k*-points in order to calculate the eigenenergies and eigenstates.[40] This grid is chosen to sample the Brillouin Zone of the periodic structure efficiently. The Hamiltonian $H(k)$ is a complex matrix whose order can vary from about 100 for a simple crystal to of order $10^5$ for a complex structure with over 1000 atoms in the unit cell. The corresponding RAM memory needs range from about 1MB to about 100GB of memory space. BLAS/LAPACK and ScaLAPACK routines are used to perform the matrix diagonalization. WIEN2k is parallelized on two levels. The first is a simple distribution of the grid of *k*-points over an array of processors, assigning a matrix $H(k)$ to each processor. This type of parallelization is analogous to the parallelization of FEFF9 and, as shown in Fig. 4, it indeed behaves similarly. We see again that a virtual cloud cluster with low-performance interconnects (slow speed, high latency) gives equal parallelization performance gains as a physical Infiniband cluster.

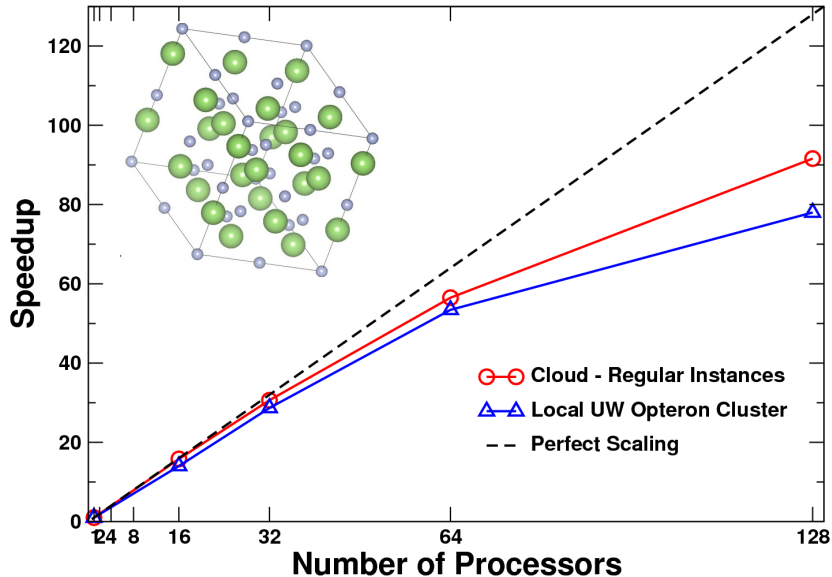

**Figure 4** Speedup of a WIEN2k calculation ("lapw1" diagonalization) of a 32-atom GaN cell on a grid of 128 *k*-points. "Regular" cloud instances with slow network connections scale just as well as a local Infiniband cluster. The inset shows the structure of the GaN cell.

The second level of parallelization in WIEN2k is of a different nature: the diagonalization of the Hamiltonian for a single *k*-point can be distributed over several processors. This is important because complex materials, characterized by a large unit cell, tend to have very large Hamiltonian matrices of order ~10,000 and larger, and a Brillouin Zone grid containing only one or very few *k*-points. ScaLAPACK routines, in conjunction with MPI, distribute the diagonalization over the processor grid. Clearly, this scheme requires the communication of large amounts of data in a time-critical way. This situation is typical of many materials science codes.

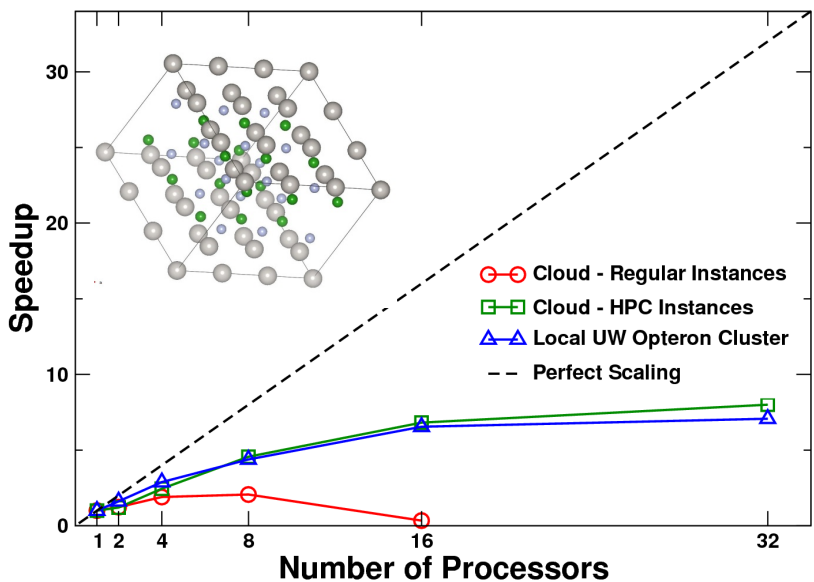

**Figure 5** Speedup of the WIEN2k calculation ("lapw1" diagonalization) of a 64-atom cell of GaN at a single *k*-point. This involves MPI/ScaLAPACK distribution of the Hamiltonian matrix across the network. The inset shows the structure of the GaN cell.

Fig. 5 shows this second type of parallelization on "Regular" EC2 cloud instances connected by the equivalent of a 100Mbps Ethernet connection. Each of these "Regular" instances has 8 cores. When the parallelization is increased beyond 8 threads, and the MPI/ScaLAPACK communication changes from intranode to internode, the calculation stalls and the speedup drops to zero, indicating network performance failure. (RAM memory is not an issue in this test.) Because of this phenomenon it is still commonly assumed in the HPC community that Cloud Computing is not suitable for scientific computing. However, when repeating the calculation on a cluster of HPC EC2 instances, which are connected by dedicated 10Gbps Ethernet network, we find that the cloud cluster can deliver the same speedup as the local Infiniband cluster. This shows that SCC is capable of serious calculations commonly associated with HPC clusters.

For a more rigorous test we calculated a much bigger system: a 1200 atom unit cell containing a slab of BN with a surface layer of Rh and a vacuum region. A few years ago this was the largest WIEN2k calculation done at that time.[43] Fig. 6 shows that once more a virtual cloud cluster of "HPC instances" delivers equal (or even slightly better) parallelization gains as the local Infiniband cluster.

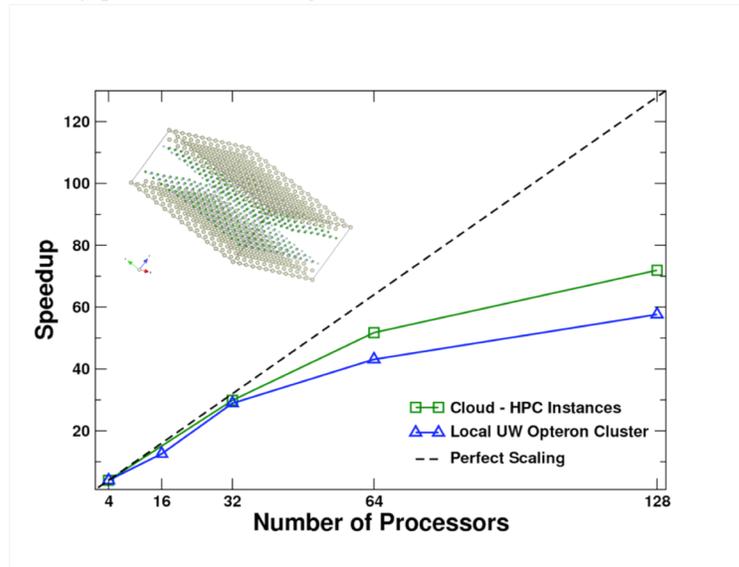

**Figure 6** Speedup for a WIEN2k calculation ("lapw1" diagonalization) of a 1200 atom cell containing a layer of Rh on a slab of BN. MPI/ScaLAPACK parallelization for a single *k*-point. The inset shows the structure of Rh@BN slab.

As an illustration we cite some absolute numbers for this calculation. Using 128 cores, the diagonalization of this complex matrix of order 56,000 took 3h48min on the local Infiniband cluster and 1h30min on a virtual HPC cloud cluster. The difference in runtime is largely due to different clock speed of 2.93-3.33GHz for the cloud cluster and 1.8GHz for the local cluster. At current (Spring 2011) EC2 pricing[44], the cost of the diagonalization was about $40, though we have not explored strategies to maximize per-dollar performance.

## 5. Integration of the SCC platform in a User Interface

### JFEFF, the GUI for FEFF on the cloud

The toolset and AMI described above constitute a complete SCC platform. However a graphical user interface is needed to create a user-friendly experience that would make SCC a convenient resource for users who are not familiar with command-line interfaces. The FEFF9 code already has a Java-based, cross-platform GUI called JFEFF.[38] JFEFF is capable of starting FEFF9 calculations either on the host computer, or on other computers accessible through ssh. We have developed an extension that links JFEFF to the SCC toolset in *NIX-like environments. A user can therefore run a FEFF9 calculation on the EC2 cloud from the comfortable GUI environment (Fig. 7) even from a laptop or a Smartphone. A full Java version of our SCC toolset will carry this functionality to all platforms and enable similar platform-independent GUIs for SCC for other materials science and quantum chemistry applications.

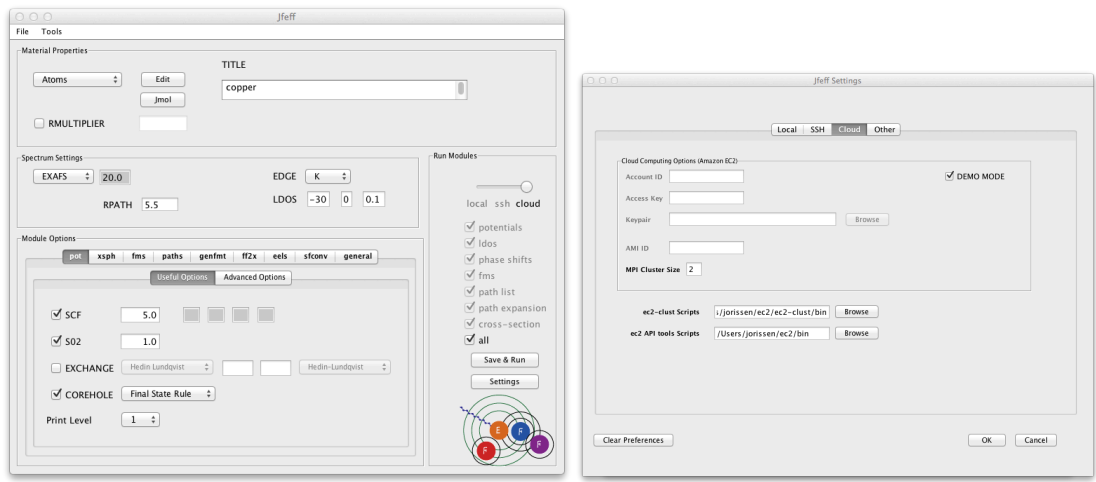

**Figure 7** Running a FEFF9 calculation on the EC2 cloud using the JFEFF GUI. Left: the main JFEFF window specifies the calculation of a spectrum of a given material (here, the K-edge EXAFS of Cu). Right: The 'Settings' window configures cloud computing, e.g. the location of EC2 login credentials, of the EC2 toolset, and the desired number of nodes in a cloud cluster. A demo mode gives users an opportunity to try a cloud calculation before they set up their own AWS account. Once the calculation is finished, JFEFF copies the output files back to the local machine. The user can then display the result on the screen (not shown).

## 6. Summary and Conclusions

We have developed a 2$^{nd}$ generation scientific AMI for the EC2 cloud, containing a number of materials-science codes and utilities that are commonly used in parallel scientific computing. Additionally, we have upgraded our Scientific Cloud Computing toolset and demonstrated large performance gains in the allocation of cloud clusters. It allows us to mount our AMI on EC2 instances with variable performance specifications. The 2$^{nd}$ generation SCC toolset is faster and more functional, without sacrificing security. Cloud clusters with hundreds of nodes can be created in reasonable setup times of a few minutes.

We have expanded our benchmarks to include scientific codes that place much higher demands on internode communication. We tested two widely used materials science codes, FEFF9 and WIEN2k. We found that cloud clusters can now provide the same speedup performance as a local Infiniband cluster. For network-heavy applications, however, it is essential to use the newly available HPC ("Cluster instance") EC2 instances, which have sufficient HPC and network capability to support network-intensive MPI/ScaLAPACK applications.

Finally, we have developed a Java-based GUI for FEFF9 on the EC2 by extending the JFEFF GUI.[38]  Thus the FEFF9 code can be run on a cloud cluster from the JFEFF GUI running on the user's local machine.  This setup could easily be used to deploy other scientific codes to the cloud as long as the requirements of data transfer to and from EC2 are modest, since AWS is currently not competitive with local solutions for the transfer and storage of very large (~>TB) data in terms of speed and cost.  Similar approaches can then be used to configure SCC AMIs for specific purposes, e.g., an AMI with applications for a theoretical x-ray beamline[45], or a quantum-chemistry AMI.

In conclusion, we have achieved the goal of developing a general Scientific Cloud Computing Platform for materials science applications that demand computational performance rather than large data handling capabilities. This platform has the potential to provide access to high performance scientific computing for general users, without requiring advanced technical computer skills or the purchase and maintenance of HPC infrastructure. The platform is also useful for developers, in that codes can be pre-installed and optimized, thus simplifying their distribution.

## Acknowledgments

The UW-SCC project is supported by NSF grant OCI-1048052. Additional EC2 cloud computer time is provided by an Amazon AWS in Education research grant PC3VBYVHQ3TASL8. The FEFF project is supported by DOE-BES grant DE-FG03-97ER45623.  We especially thank AWS and in particular Deepak Singh for support and encouragement. We also thank Jeff Gardner for valuable discussions and comments.